\documentclass[a4paper,12pt]{article}

\usepackage[english]{babel}
\usepackage[utf8x]{inputenc}
\usepackage[T1]{fontenc}
\usepackage{url}
\usepackage{authblk}
\usepackage{newtxtext,newtxmath} % Times New Roman font

\usepackage{geometry}
\usepackage{setspace}
\usepackage[numbers,sort&compress]{natbib}

\usepackage{graphicx}
\usepackage{caption}
\DeclareCaptionLabelSeparator{bar}{\,$\mid$\,}
\usepackage{amsmath} 
\usepackage{amssymb} 
\usepackage[detect-all]{siunitx}

\newcommand*\samethanks[1][\value{footnote}]{\footnotemark[#1]}

\usepackage{color}

\begin{document}
\renewcommand{\abstractname}{\vspace{-\baselineskip}}

\title{Field-free deterministic ultra fast creation of skyrmions by spin orbit torques} %Title of paper
\author[1]{Felix B\"{u}ttner \thanks{Corresponding author: \texttt{felixbuettner@gmail.com}} \footnote{These authors contributed equally to this work} }
\author[1]{Ivan Lemesh \samethanks}
\author[2]{Michael Schneider}
\author[2]{Bastian Pfau}
\author[3]{Christian M. G\"unther}
\author[2]{Piet Hessing}
\author[2]{Jan Geilhufe}
\author[1]{Lucas Caretta}
\author[2]{Dieter Engel}
\author[4]{Benjamin Kr\"uger}
\author[5]{Jens Viefhaus}
\author[2,3]{Stefan Eisebitt}
\author[1]{Geoffrey S. D. Beach}
\affil[1]{Department of Materials Science and Engineering, Massachusetts Institute of Technology, Cambridge, Massachusetts 02139, USA}
\affil[2]{Max-Born-Institut, Max-Born-Stra\ss e 2A, 12489 Berlin, Germany}
\affil[3]{Institut f\"ur Optik und Atomare Physik, Technische Universit\"at Berlin, Hardenbergstra\ss e 36, 10623 Berlin, Germany}
\affil[4]{Institute for Laser Technologies in Medicine and Metrology at the University of Ulm, Ulm, Germany}
\affil[5]{Deutsches Elektronen-Synchrotron (DESY), FS-PE, Notkestra{\ss}e 85, 22607 Hamburg, Germany}

\date{\today}

\maketitle 

\clearpage

\textbf{Magnetic skyrmions are currently the most promising option to realize current-driven magnetic shift registers \cite{fert_skyrmions_2013,wiesendanger_nanoscale_2016,rosch_skyrmions:_2013}. A variety of concepts to create skyrmions were proposed and demonstrated \cite{sampaio_nucleation_2013,romming_writing_2013,koshibae_memory_2015,iwasaki_current-induced_2013,woo_observation_2016,zhou_reversible_2014,jiang_blowing_2015,hrabec_current-induced_2016,legrand_room-temperature_2017,schott_skyrmion_2016}. However, none of the reported experiments show controlled creation of single skyrmions using integrated designs. Here, we demonstrate that skyrmions can be generated deterministically on subnanosecond timescales in magnetic racetracks at artificial or natural defects using spin orbit torque (SOT) pulses. The mechanism is largely similar to SOT--induced switching of uniformly magnetized elements \cite{liu_spin-torque_2012,garello_ultrafast_2014,lee_threshold_2013}, but due to the effect of the Dzyaloshinskii-Moriya interaction (DMI), external fields are not required. Our observations provide a simple and reliable means for skyrmion writing that can be readily integrated into racetrack devices.}

Magnetic skyrmions are small particle-like domains in an out-of-plane magnetized film, envisioned as information carriers in racetrack devices in which they can be created, deleted, and shifted by current \cite{fert_skyrmions_2013,wiesendanger_nanoscale_2016,rosch_skyrmions:_2013}. Their defining property is a spherical topology, manifesting in a defect-free domain wall \cite{nagaosa_topological_2013,hellman_interface-induced_2016}. Skyrmions can be found in a variety of systems, where they are stabilized by a combination of stray field energies, external magnetic fields, and anisotropic exchange interactions \cite{jiang_blowing_2015,buttner_dynamics_2015,moreau-luchaire_additive_2016,nagaosa_topological_2013,woo_observation_2016,yu_variation_2016}. The anisotropic exchange-like DMI favors homochiral and defect-free skyrmionic spin structures. SOT--driven motion of such homochiral skyrmions is deterministic, rendering systems with large DMI most relevant for applications. Asymmetric multilayers of non-magnetic heavy metals with strong spin orbit interactions and transition metal ferromagnetic layers provide large and tunable DMI \cite{litzius_skyrmion_2016,moreau-luchaire_additive_2016,woo_observation_2016,jiang_direct_2016,jiang_blowing_2015}. Also, the non-magnetic heavy metal layer can serve to inject a vertical spin current with transverse spin polarization into the ferromagnetic layer through the spin Hall effect (SHE) \cite{sinova_spin_2015}. The injected angular momentum leads to two torques on the local magnetization, a damping-like (DL) and a field-like (FL) torque \cite{brataas_spin-orbit_2014}, both of which pull the magnetization into the film plane. These SOTs can be used to completely switch the magnetization in out-of-plane magnetized ferromagnetic elements, but the switching is deterministic only in the presence of a symmetry-breaking in-plane field \cite{liu_spin-torque_2012,garello_ultrafast_2014,lee_threshold_2013}. SOT has also been shown to lead to domain nucleation in continuous films \cite{huang_initialization-free_2017} and to stochastic nucleation of skyrmions in magnetic tracks \cite{legrand_room-temperature_2017}. However, no practical means to controllably create individual skyrmions in an integrated device design has yet been reported. Here, we demonstrate that sub-nanosecond SOT pulses can deterministically generate single skyrmions at custom defined positions in a magnetic racetrack, using the same current path that is used for the shifting operation.  The implementation is most simple: a defect such as a constriction in the magnetic track, which is easily fabricated in the same process as the track itself, can serve as a skyrmion generator. Hence, the concept is applicable to any track geometry, including three-dimensional designs \cite{parkin_magnetic_2008}, providing an integrated solution to the writing problem in skyrmion-based devices.

\begin{figure}
\centering
\includegraphics[width=\textwidth]{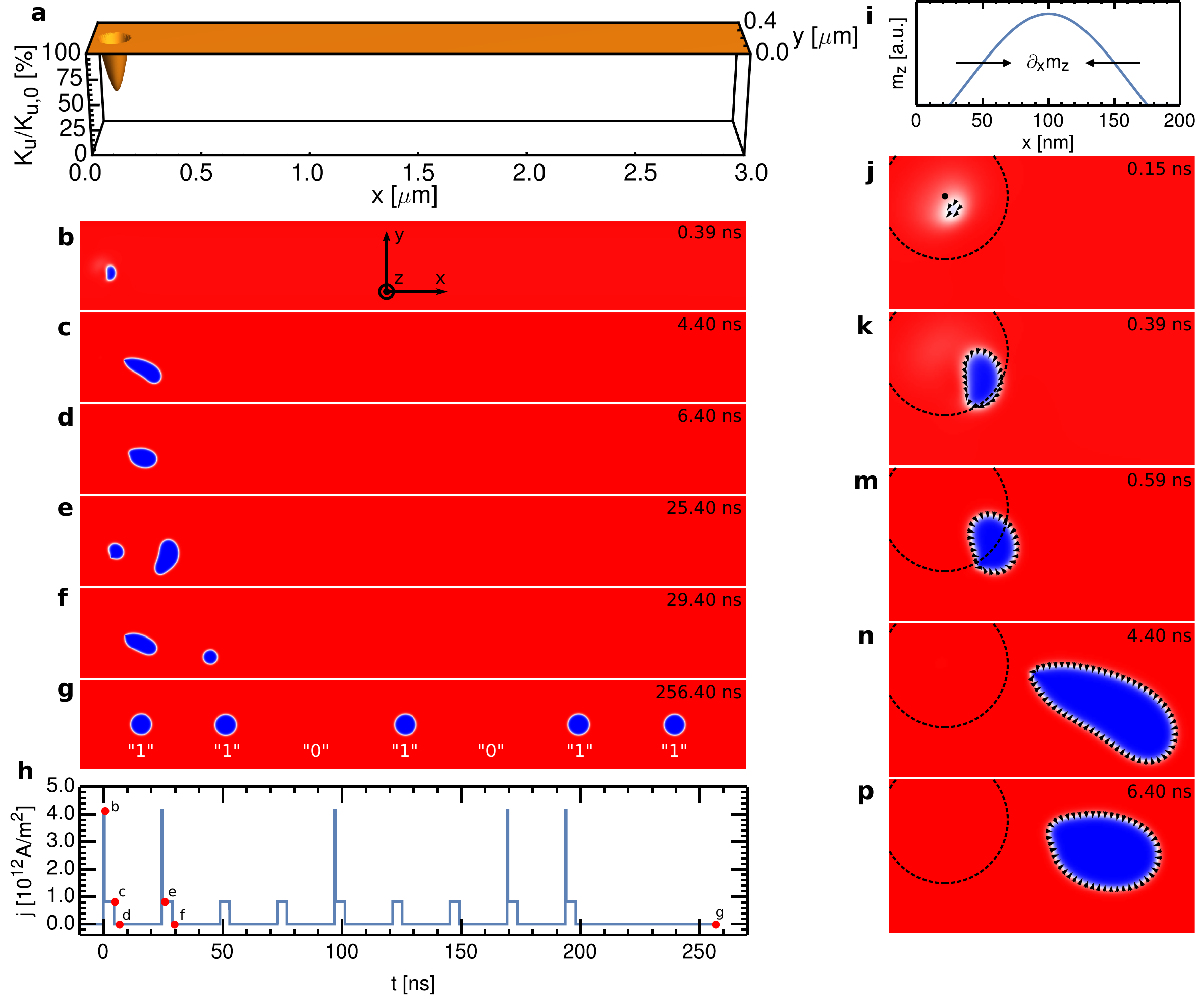}
\caption{\textbf{Simulation of a bit sequence in a racetrack memory created by SOT skyrmion generation near a pinning center.} \textbf{a}, Anisotropy distribution in the $3\times\SI{0.4}{\micro m^2}$ magnetic strip. The area of reduced anisotropy has a diameter of \SI{150}{nm} and constitutes a pinning site, as typically found in sputtered multilayers. \textbf{b--g}, Temporal evolution of the magnetization in the strip following the injection of SOT pulses. The color decodes $m_z$ (blue: $m_z>0$, red: $m_z<0$). The injected pulse pattern is plotted in \textbf{h}, where also the time corresponding to the snapshots \textbf{b--g} is indicated. The width of the write pulses is \SI{0.4}{ns}, the width of the shift pulses is \SI{4}{ns}, and the relaxation in between lasts for \SI{20}{ns}. Large pulse amplitudes are required because of a conservatively chosen value for the Gilbert damping of $\alpha=0.5$ and the absence of temperature effects. The final skyrmion pattern in panel \textbf{g} corresponds to the bit sequence ``1101011'', with the presence (absence) of a skyrmion representing a logical ``1'' (``0'') as depicted in the figure. \textbf{i}, Schematic cross section of $m_z$ across the defect at the start of the SOT pulse (before the domain nucleation). The arrows indicate the sign of the gradient of $m_z$, pointing towards the center of the defect. \textbf{j--p}, Zoom-in of the defect area. The defect circumference is sketched in each image by a dashed circle. The center is marked with a black dot in \textbf{j}. The times in the images correspond to the pulse shape in \textbf{h}.
}
\label{fig:1}
\end{figure}

The concept of SOT--induced skyrmion nucleation is illustrated in the micromagnetic simulations in Fig.~\ref{fig:1}, using the example of a pinning site (area of reduced anisotropy) in a magnetic nanotrack.  The track is initially magnetized along the negative $z$ direction ($m_z<0$) and a charge current pulse is injected along the $+x$ direction.  The SHE leads to injection of $\mathbf{p}=-\hat{y}$ polarized magnetic moments into the ferromagnet, generating DL and FL SOTs (see Methods). It is well known that SOT can lead to the switching of a ferromagnet if a longitudinal field is applied \cite{liu_spin-torque_2012,garello_ultrafast_2014,lee_threshold_2013}, see also Supplementary Information. Here, we observe local switching even without any in-plane fields applied, which becomes possible by combining DMI with a non-uniform out-of-plane magnetization. Due to the reduced anisotropy, the SOT--induced reduction of $|m_z|$ is largest inside the defect (Fig.~\ref{fig:1}i), thereby generating a gradient $\nabla m_z$ pointing towards the center of the defect. The interfacial DMI that prefers left-handed domain walls (negative $D_i$ in our convention) transforms this gradient into a field $\mathbf{H}_\text{DMI}=\frac{(-2D_i)}{M_s}\left(\partial_xm_z,\partial_ym_z,-\partial_xm_x-\partial_ym_y\right)$ that points towards the defect. At the bottom-right side of the defect, the DMI field has a negative $x$ component as required for switching and a positive $y$ component that compensates the field from the FL SOT. Hence, this is where $m_z$ switches sign. Once a reversed domain has formed, the moments in the domain wall align with the injected moments $\mathbf{p}$ wherever possible due to the FL SOT. This can be observed in Fig.~\ref{fig:1}j, where we also see that the new domain is non-topological because that would require all magnetization orientations to be present (and $+y$ oriented moments are missing). The topological transition happens at a later point, as we will see now.

In the simulations, we use a strong and short pulse to nucleate a reverse domain (write pulse) and a longer but weaker pulse to drive it away from the defect (shift pulse), see Fig.~\ref{fig:1}h. As illustrated in Figs.~\ref{fig:1}m--p, the weak pulse allows the magnetization in the domain wall to relax to the preferred left-handed (inwards-pointing) chirality almost everywhere apart from a localized pair of vertical Bloch lines (VBLs). Due to the VBLs, the domain is still non-topological. Subsequently, the left-handed N\'eel domain wall moves with the current while the VBLs remain nearly fixed in position. The center of the domain, however, moves significantly away from the pinning center. Once the current is switched off, the energetically unfavorable VBLs are expelled and the domain transforms into a topological skymion (Fig.~\ref{fig:1}p). In Figs.~\ref{fig:1}b--h we demonstrate that an evenly spaced bit sequence, where logical 1 and 0 are represented by the presence or absence of a skyrmion, respectively, can be obtained by repeatedly injecting write-shift or shift-only current pulses. Note that skyrmions do not reverse to non-topological domains even under the strong write current pulses. Skyrmion trajectories have a negative $y$ component due to the skyrmion Hall effect (Fig.~\ref{fig:1}f) \cite{litzius_skyrmion_2016,jiang_direct_2016}. Therefore, a \SI{20}{ns} pause was included after each shift pulse to allow the skyrmions return to the center of the wire. 

\begin{figure}
\centering
\includegraphics[width=\textwidth]{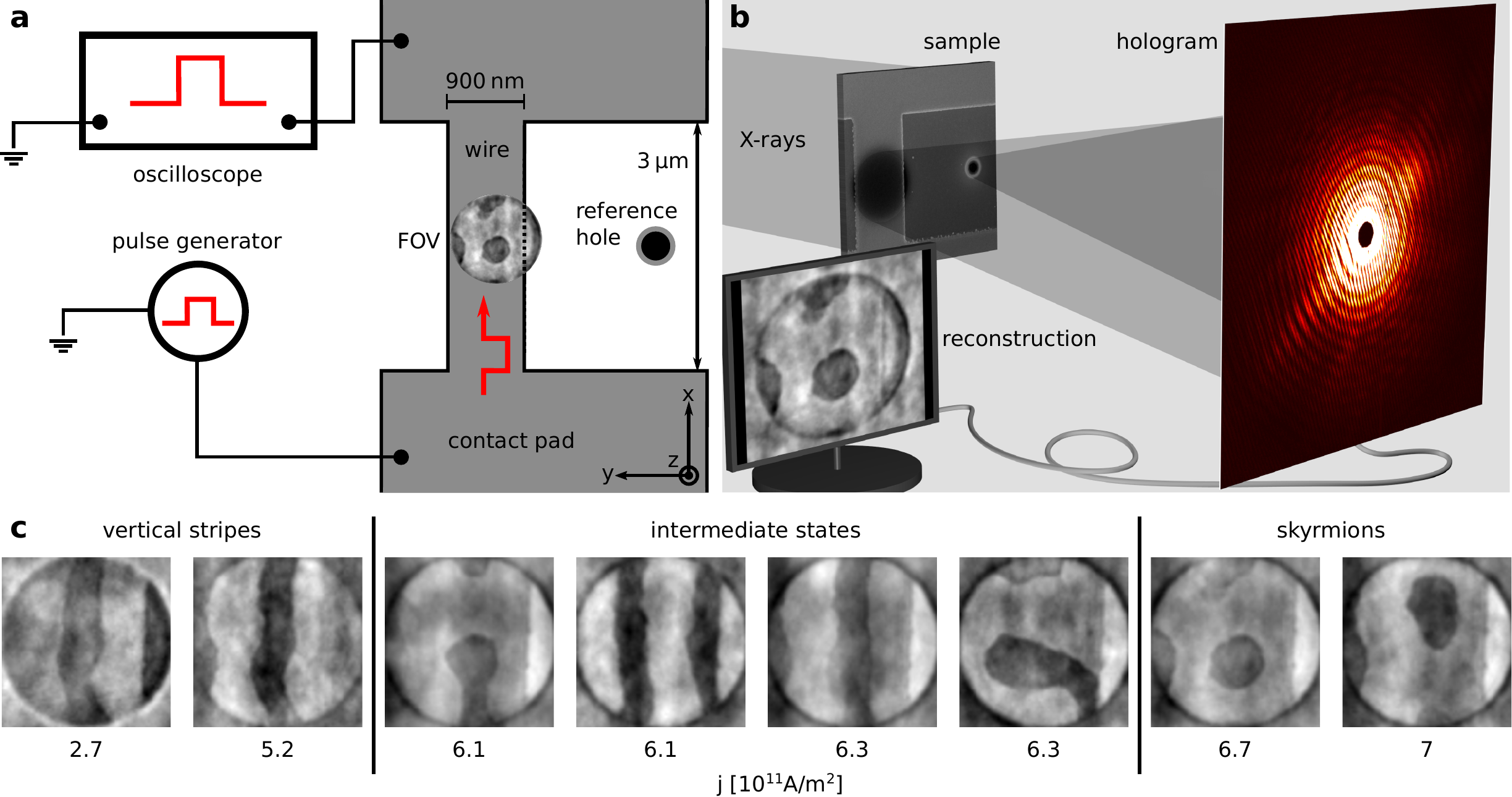}
\caption{\textbf{Experimental setup and the effect of pulse trains.} \textbf{a}, Schematic top-view of the sample and connection to the pulse electronics. The sample consists of an H-shaped magnetic material, the central part of which is the magnetic wire. The wire is \SI{3}{\micro m} long and \SI{900}{nm} wide. The center of the wire is imaged holographically (see \textbf{b}) with a circular field of view (FOV). An example magnetic texture in the FOV is overlaid. Part of the FOV is outside of the wire, which makes it easy to recognize the image orientation. The wider parts of the magnetic material are connected to gold contact pads and from there to a pulse generator and an oscilloscope. \textbf{b}, Imaging setup. The magnetic wire was grown on a transparent Si$_3$N$_4$ membrane. On the back side of the membrane, two holes were prepared in an opaque gold film. The larger one (\SI{1}{\micro m} in diameter) is aligned to the wire and defines the FOV. The second hole is point-like (with a diameter of typically a few tens of nanometers) and has a distance of \SI{3}{\micro m} to the larger hole. It defines the reference beam for the holographic imaging. The structure is exposed to circularly polarized X-rays. The interference of the transmitted beam of the two holes forms the hologram, from which an image of the out-of-plane magnetization in the wire can be reconstructed. \textbf{c}, Nucleation of skyrmions and stripes with unipolar pulse trains. The injection of millions of pulses (each \SI{180}{ns} in width) leads to the formation of stripe domains at low current densities. The stripes are aligned parallel to the injected current. At large current densities, skyrmions prevail.
}
\label{fig:2}
\end{figure}

We now verify the predictions of the simulations experimentally using X-ray holography \cite{eisebitt_lensless_2004,buttner_dynamics_2015,buttner_dynamic_2017}. We use a multilayer stack consisting of 15 repeats of Pt(\SI{2.7}{nm})/Co$_{60}$Fe$_{20}$B$_{20}$(\SI{0.8}{nm})/MgO(\SI{1.5}{nm}) grown on a Si$_3$N$_4$ membrane with a seed layer of Ta(\SI{2.3}{nm})/Pt(\SI{3.7}{nm}). Such multilayers have previously been shown to host left-handed N\'eel skyrmions due to the large DMI at the Pt/CoFeB interface \cite{litzius_skyrmion_2016,woo_observation_2016}. The Pt also serves as a SOT source. The film was patterned into a \SI{3}{\micro m} long and \SI{900}{nm} wide contacted track (Figs.~\ref{fig:2}a,b). On the back side of the membrane, two holes were prepared in an opaque Au film, the larger one with \SI{1}{\micro m} diameter defining the field of view (FOV) and the smaller one providing the reference beam for X-ray holographic imaging \cite{eisebitt_lensless_2004,buttner_dynamic_2017}, allowing us to reconstruct images of the distribution of $m_z$ within the FOV. The magnetic state was imaged after a sequence of (i) saturating the film with a large $z$-axis bias field, (ii) reducing the field to a value where domains can exist but do not nucleate spontaneously, and (iii) injecting single rectangular-shaped SOT current pulses or unipolar pulse trains. In Fig.~\ref{fig:2}c we show the effect of injecting millions of current pulses before taking the image. The result depends on the current amplitude: At low current densities, stripe domains appear and the stripes are consistently oriented parallel to the injected current. In applications, these low current densities should hence be avoided. At high current densities, skyrmions prevail. The threshold in our material is approximately \SI{6.5e11}{A/m^2}.

\begin{figure}
\centering
\includegraphics[width=\textwidth]{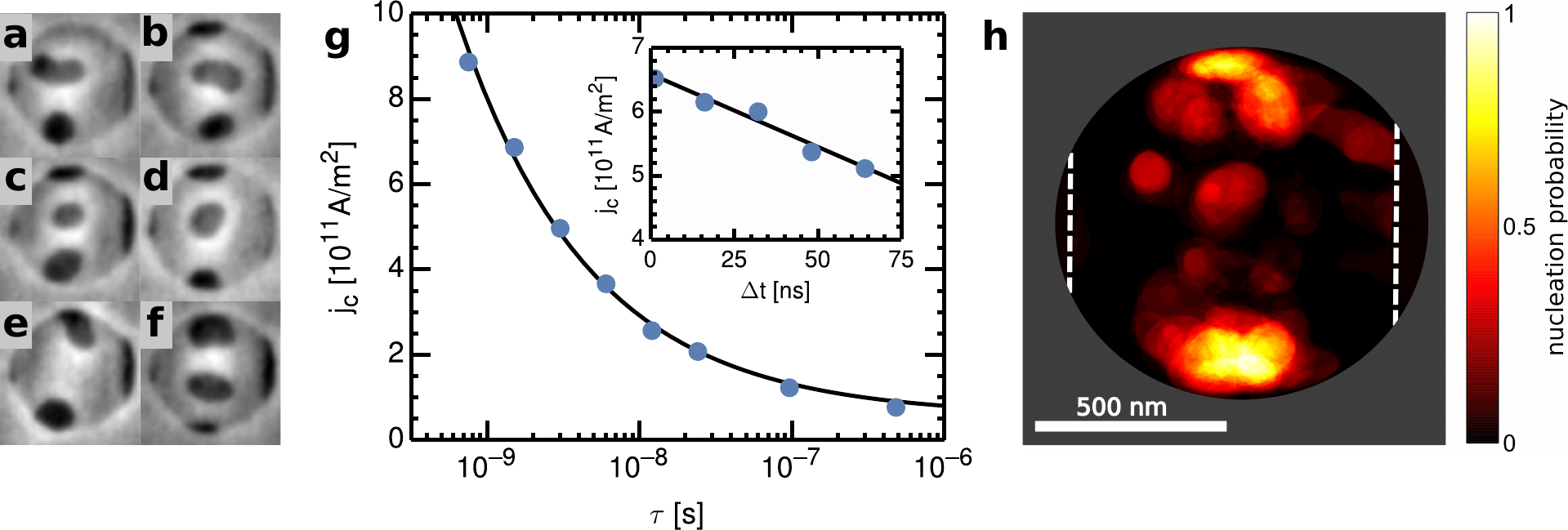}
\caption{\textbf{Creation of skyrmions at natural defects.} \textbf{a--f}, Examples of skyrmions nucleated with single \SI{12}{ns} long pulses of \SI{2.6e11}{A/m^2} amplitude. Before each image, a large magnetic field was temporarily applied to saturate the sample. Skyrmions are often slightly elongated in $y$ direction. The black contrast at the right side of the images is the edge of the wire and of non-magnetic origin. \textbf{g}, Critical current density $j_c$ to nucleate domains with a single pulse as a function of pulse width $\tau$. Inset: Critical current density $j_c$ for reproducible domain nucleation with a \SI{96}{ns} long base pulse of \SI{0.95e11}{A/m^2} and a variable amplitude \SI{1.5}{ns} peak pulse of variable delay $\Delta t$ with respect to the start of the base pulse. The base pulse amplitude is \SI{25}{\percent} below the threshold for skyrmion nucleation, hence serving here primarily as a Joule heater: The longer the delay of the peak pulse, the warmer is the wire and the critical current density decreases. The lines are guides to the eye. \textbf{h}, Spatial probability distribution of finding a reversed magnetization after the application of single pulses in the experimental data. Areas of high nucleation probability correspond to strong pinning centers according to the simulation results. The white dashed lines mark the edges of the track.
}
\label{fig:3}
\end{figure}

The effect of single SOT pulses is studied systematically in Fig.~\ref{fig:3} as a function of pulse width $\tau$ and amplitude $j$. We find that reversed domains are generated with certainty for amplitudes above a sharp threshold $j_c(\tau)$. In Figs.~\ref{fig:3}a--f we show a selection of skyrmions generated by $\tau=\SI{12}{ns}$ and $j=\SI{2.6e11}{A/m^2}$ pulses. We find skyrmions in all images. The same effect is observed for larger currents, at least up to \SI{6e11}{A/m^2}. Already at slightly lower amplitudes of \SI{2.2e11}{A/m^2}, however, the probability of observing a skyrmion reduces to approximately \num{1/3}. The functional dependence $j_c(\tau)$, depicted in Fig.~\ref{fig:3}g, is qualitatively consistent with the established model for SOT switching \cite{garello_ultrafast_2014}: At small $\tau$, $j_c$ drops rapidly because a minimum time-integrated injected angular momentum is required for switching. This effect saturates due to Gilbert damping at the current density required for static switching (see Supplementary Information). At longer pulses, thermal effects help to overcome the energy barrier and $j_c(\tau)$ follows an Arrhenius law. Quantitatively, the expected $1/\tau$ and $1/\log(\tau)$ functions \cite{garello_ultrafast_2014} are not able to describe our data. We suggest that the reason for this deviation is that the temperature is not constant due to Joule heating. We confirm this experimentally by using composite pulses consisting of a short pulse on top of a long, low amplitude pulse. The pulses start with a relative delay of $\Delta t$. Switching happens within the fixed time scale of the short pulse. Assuming constant temperature, we therefore expect that $j_c$ does not depend on $\Delta t$. In contrast, we observe that $j_c$ is significantly reduced towards the end of the long pulse (inset of Fig.~\ref{fig:3}g). This effect can be explained by thermally assisted switching and a still-increasing temperature \SI{70}{ns} after the start of the long pulse. The finite temperature effects (as well as possibly a lower damping than assumed in the simulations) also explain why our experimentally determined current densities are much lower than in the zero temperature simulations.

Skyrmions can be shifted along the wire by applying unipolar sub-threshold current pulses (pulses that move skyrmions but do not create new ones), see Supplementary Movie. Skyrmions move from pinning site to pinning site, as evidenced by the similar skyrmion positions observed in various images. Skyrmions move in the direction of current (against the electron flow), which finally confirms their left-handed chirality, the consequently defect free domain walls, and thus the single spherical topology (topological charge $N=1$) \cite{litzius_skyrmion_2016}. The maximum displacement with 10 pulses of \SI{6}{ns} length is \SI{500}{nm}, corresponding to a velocity of \SI{8}{m/s}, which is in agreement with previous observations \cite{woo_observation_2016}. The skyrmion trajectories are mostly parallel to the current, i.e., showing no skyrmion Hall effect, which is expected considering that the current density of $j=\SI{2.2e11}{A/m^2}$ is only sightly larger than the minimum amplitude to induce motion in this material \cite{jiang_direct_2016}. After 230 pulses all skyrmions have disappeared from the field of view and none appear following more pulses because skyrmions were generated only within the wire and the wire has a finite length.

The spatial probability distribution of finding negative $m_z$ (skyrmions or domains) after the application of a single pulse is depicted in Fig.~\ref{fig:3}h. The discrete hot spots in this distribution confirm localized defects to be the origin of skyrmion nucleation. Skyrmions are mostly found within the wire, detached from the edge, which is important for applications. Short pulses predominantly create round skyrmions and -- rarely -- domains attached to the wire edges. At long pulses, in contrast, we frequently encounter stripe domains oriented parallel to the current, as also observed in Fig.~\ref{fig:2}a. Short and strong pulses are hence preferable for skyrmion nucleation and for skyrmion motion, which is also helpful for high speed operations.

\begin{figure}
\centering
\includegraphics[width=\textwidth]{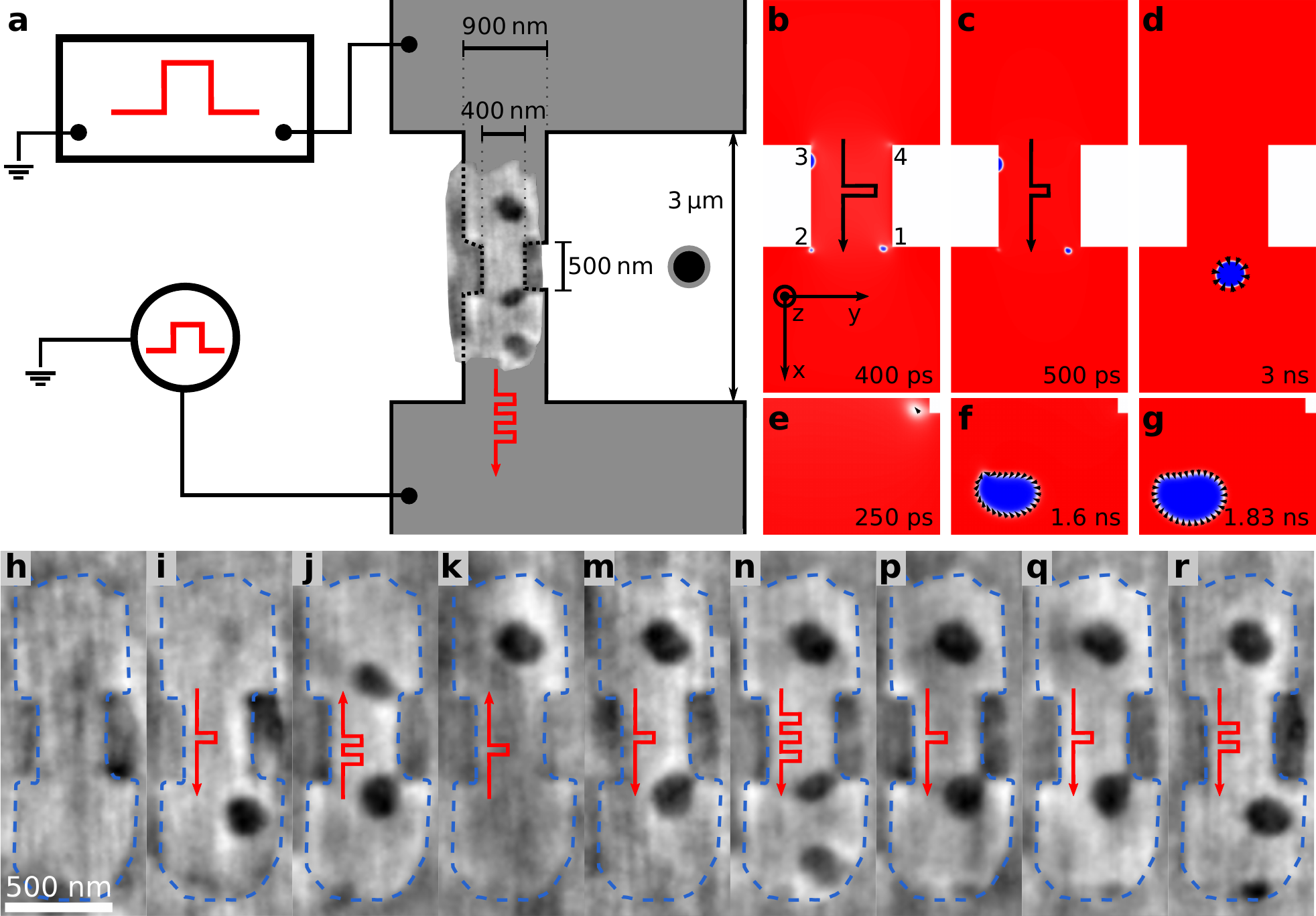}
\caption{\textbf{Demonstration of single skyrmion generation and subsequent motion.} \textbf{a}, Schematic view of the new sample design. With respect to Fig.~\ref{fig:2}a, a constriction has been added to the wire and the field of view has been enlarged. \textbf{b--g}, Micromagnetic simulations of the nucleation process at the constriction, using again a nucleation pulse ($j=\SI{7.0e12}{A/m^2}$, $\tau=\SI{0.4}{ns}$) followed by a shift pulse  ($j=\SI{3.5e12}{A/m^2}$, $\tau=\SI{1.2}{ns}$). Current densities refer to the center of the constriction. The time corresponding to each image is denoted in the bottom-right. The corners are numbered in \textbf{b}. \textbf{e--g}, Magnification of corner 1. \textbf{h--r}, Experimental confirmation of controlled skyrmion nucleation near a constriction. The blue dotted line is the border of the constricted wire within the field of view. The red arrows indicate the direction and number of single $j=\SI{5.8e11}{A/m^2}$, $\tau=\SI{6}{ns}$ pulses applied prior to imaging the respective frame.
}
\label{fig:4}
\end{figure}

Our results so far demonstrate reproducible skyrmion generation and explain the underlying mechanism. However, the location of the nucleated skyrmions was not fully predictable, as required for applications, since here we rely upon naturally-occurring defects. To deterministically generate skyrmions at a predetermined location we use a tailored defect -- namely a constriction -- instead of naturally present pinning sites. We cut the constriction from an existing wire using focused ion beam milling and also enlarge the field of view, as depicted in Fig.~\ref{fig:4}a. Simulations suggest that domains nucleate at three out of four corners of the constriction, but only one of those domains is stable after switching off the pulse (Figs.~\ref{fig:4}b--g). The difference between corner 1 and corner 2 is mainly that the skyrmion in corner 1 detaches more easily from the edge because of the skyrmion Hall effect. Once it is sufficiently far away from the edges it becomes more stable, which explains why this domain survives and the other two annihilate. Note, however, that a different shape of the corners can lead to nucleation of a stable domain at corner 2 instead of corner 1. Importantly, though, domains leave the constriction exclusively in the direction of the applied current, independent of pulse shape and details of the geometry. The nucleation mechanism is the same as at bulk pinning sites, see Figs.~\ref{fig:4}e--g. The gradient of $m_z$ comes from the increased current density at the corners and tilting of spins at the boundaries due to DMI \cite{rohart_skyrmion_2013,iwasaki_current-induced_2013}.

The experiments confirm the predictions of the micromagnetic simulations (Figs.~\ref{fig:4}h--r). We start with a saturated wire (Fig.~\ref{fig:4}h). Subsequently, we record a sequence of images at constant external field, i.e., without saturating in between. Before each image was acquired, we injected unipolar SOT current pulses with a magnitude below the nucleation threshold for the wide part of the wire. The number of pulses and their direction is indicated by the red pulse shapes in the images. We observe reproducible nucleation of single skyrmions selectively at one of the exits of the constriction, determined by the direction of the applied current. Following their nucleation, skyrmions can be moved within the wide part of the wire by applying more current pulses. New skyrmions are generated once the existing skyrmion has moved away and sufficient space is available. Note that the constriction itself is too narrow for skyrmions to survive. Hence, skyrmions are repelled from the entrance of the constriction and annihilate when they are forced to enter nonetheless (Figs.~\ref{fig:4}j,k). Our sample shows significant natural pinning as revealed by the resistance to motion of the nucleated skyrmions. However, this can be amended by further material and fabrication process optimization \cite{litzius_skyrmion_2016}.

Our results demonstrate that skyrmion generators can be integrated into racetrack devices in the most simple way: by patterning of notches or constrictions. We thereby confirm earlier theoretical concepts \cite{iwasaki_current-induced_2013} and extend them to SOT devices and materials with interfacial DMI. These simple generators provide full control over where and when skyrmions are nucleated at time scales and voltages matching present-day computer architectures. Constriction-type skyrmion generators can become transparent by making them wide enough to comfortably fit a skyrmion, which is important if such elements are to be used in positions other than the ends of a racetrack memory. All ambiguity regarding the exact point of nucleation can be removed by using triangular-shaped notches with only one corner. Hence, the last remaining fundamental challenge for a prototype skyrmion racetrack memory is the controlled annihilation of a skyrmion.

\section*{Acknowledgements}

This work was supported by the U.S. Department of Energy (DOE), Office of Science, Basic Energy Sciences (BES) under Award \#DE-SC0012371. FB acknowledges financial support by the German Science Foundation under grant number BU 3297/1-1. 

% \section*{Authors contributions}
% 
% FB, BP, SE, and GSDB conceived and designed the experiment. FB, IL, MS, CMG, and DE prepared and pre-characterized the samples. FB, IL, MS, BP, PH, JG, and LC performed the experiments with support by JV. BP and PH reconstructed the holographic images. IL, FB, and BK performed the micromagnetic simulations. FB drafted the manuscript. SE and GSDB supervised the project.  All authors discussed the results, the implications, and the figures and commented on the manuscript.
% 
% \section*{Additional information}
% FB and IL contributed equally to this work. Supplementary information and videos of all simulations are available in the online version of the paper. Reprints and
% permissions information is available online at www.nature.com/reprints.
% Correspondence and requests for materials should be addressed to FB (felixbuettner@gmail.com).
% 
% \section*{Competing financial interests}
% The authors declare no competing financial interests.

\section*{Methods summary}

\subsection*{Simulation details}

In the simulation images, blue indicates positive $m_z$ and red indicates negative $m_z$. Simulations were performed at zero temperature using the MuMax software \cite{vansteenkiste_design_2014}, employing the effective medium model \cite{woo_observation_2016} to simulate the experimental magnetic multilayer as one isotropic medium with one cell in $z$ direction. Material parameters were chosen to mimic the materials in the experiments. Specifically, the single layer saturation magnetization $M_s=\SI{1.12e6}{A/m}$ and anisotropy constant $K_u=\SI{9.78e5}{J/m^3}$ were determined experimentally by vibrating sample magnetometry. Exchange and DMI constants $A=\SI{e-11}{J/m}$ and $D_i=\SI{1.5e-3}{J/m^2}$ were chosen in agreement with the observed stripe domain width of the demagnetized state of \SIrange{150}{200}{nm}, as measured by magnetic force microscopy (and confirmed in Fig.~\ref{fig:2}c). Simulation cells with a lateral edge length of \SI{2.25}{nm} were used. Out-of-plane magnetic fields of $\mu_0H_z=-\SI{57.5}{mT}$ and $-\SI{50}{mT}$ were applied in Figs.~\ref{fig:1} and \ref{fig:4}, respectively. SOT effects were simulated using the available solver for Slonczewski spin values. To convert the parameters of this solver to transverse spin Hall currents, a spin Hall angle of $\theta_\text{SH}=0.07$ was assumed. The coefficient of the FL torque $b_j$ was adjusted to half of the coefficient of the DL torque $a_j$. The current distribution for the constricted wire (Figs.~\ref{fig:4}b--g) was determined using a finite differences solver with a given constant voltage at the sample edges. 

\subsection*{Experimental details}

Imaging was performed at beamline P04 at PETRA III, DESY, Germany, using X-ray holography with circular polarized X-rays of \SI{778}{eV} photon energy \cite{buttner_dynamic_2017}. The reference holes had a diameter of \SI{25}{nm}, \SI{120}{nm}, and \SI{110}{nm} in Figs.~\ref{fig:2}, \ref{fig:3}, and \ref{fig:4}, respectively. Samples were investigated in vacuum at room temperature. Purely out-of-plane oriented fields of  $\mu_0H_z=-\SI{3.2}{mT},$ $-\SI{6.1}{mT}$, and $-\SI{7.1}{mT}$ were applied during the imaging of the magnetization in Figs.~\ref{fig:2}, \ref{fig:3}, and \ref{fig:4}, respectively.
In all experimental magnetization images, black indicates positive $m_z$ and white indicates negative $m_z$.

% \clearpage

% \bibliographystyle{naturemag_noURL}
% \bibliography{references}

\end{document}